%% file: main.tex
\documentclass[twocolumn,tighten]{aastex63}

\usepackage{bm}

\newcommand\kms{\ifmmode{\rm km\thinspace s^{-1}}\else km\thinspace s$^{-1}$\fi}
\newcommand\ms{\ifmmode{\rm m\thinspace s^{-1}}\else m\thinspace s$^{-1}$\fi}

\newcommand{\tw}{TWA~3}
\newcommand{\twelve}{${}^{12}$CO}
\newcommand{\thirteen}{${}^{13}$CO}

\newcommand{\imut}{\ensuremath{\theta}}

\begin{document}

\title{A coplanar circumbinary protoplanetary disk in the TWA 3 triple M dwarf system}

\correspondingauthor{Ian Czekala}
\email{iczekala@berkeley.edu}

\author[0000-0002-1483-8811]{Ian Czekala}
\altaffiliation{NASA Hubble Fellowship Program Sagan Fellow}
\affiliation{Department of Astronomy and Astrophysics, 525 Davey Laboratory, The Pennsylvania State University, University Park, PA 16802, USA}
\affiliation{Center for Exoplanets and Habitable Worlds, 525 Davey Laboratory, The Pennsylvania State University, University Park, PA 16802, USA}
\affiliation{Center for Astrostatistics, 525 Davey Laboratory, The Pennsylvania State University, University Park, PA 16802, USA}
\affiliation{Institute for Computational \& Data Sciences, The Pennsylvania State University, University Park, PA 16802, USA}
\affiliation{Department of Astronomy, 501 Campbell Hall, University of California, Berkeley, CA 94720-3411, USA}

\author[0000-0003-3133-3580]{\'Alvaro Ribas}
\affiliation{European Southern Observatory (ESO), Alonso de C\'ordova 3107, Vitacura, Casilla 19001, Santiago de Chile, Chile}

\author[0000-0003-3713-8073]{Nicol\'as Cuello}
\affiliation{Univ. Grenoble Alpes, CNRS, IPAG (UMR 5274), F-38000 Grenoble, France}

\author[0000-0002-6246-2310]{Eugene Chiang}
\affiliation{Department of Astronomy, 501 Campbell Hall, University of California, Berkeley, CA 94720-3411, USA}
\affiliation{Department of Earth and Planetary Science, University of California at Berkeley, Berkeley, CA 94720-4767, USA}

\author[0000-0003-1283-6262]{Enrique Mac\'ias}
\affiliation{Joint ALMA Observatory, Alonso de C\'ordova 3107, Vitacura, Santiago, Chile}
\affiliation{European Southern Observatory (ESO), Alonso de C\'ordova 3107, Vitacura, Casilla 19001, Santiago de Chile, Chile}

\author[0000-0002-5092-6464]{Gaspard Duch\^{e}ne}
\affiliation{Department of Astronomy, 501 Campbell Hall, University of California, Berkeley, CA 94720-3411, USA}
\affiliation{Univ. Grenoble Alpes, CNRS, IPAG (UMR 5274), F-38000 Grenoble, France}

\author[0000-0003-2253-2270]{Sean M. Andrews}
\affiliation{Center for Astrophysics \textbar\ Harvard \& Smithsonian, 60 Garden Street, Cambridge, MA 02138, USA}

\author[0000-0001-9227-5949]{Catherine C. Espaillat}
\affiliation{Department of Astronomy, Boston University, 725 Commonwealth Avenue, Boston, MA 02215, USA}

\begin{abstract}
We present sensitive ALMA observations of TWA~3, a nearby, young ($\sim$10\,Myr) hierarchical system composed of three pre-main sequence M3--M4.5 stars. For the first time, we detected \twelve\ and \thirteen\ $J$=2-1 emission from the circumbinary protoplanetary disk around TWA~3A. We jointly fit the protoplanetary disk velocity field, stellar astrometric positions, and stellar radial velocities to infer the architecture of the system. The Aa and Ab stars ($0.29\pm0.01\,M_\odot$ and $0.24\pm0.01\,M_\odot$, respectively) comprising the tight ($P=35$\,days) eccentric ($e=0.63\pm0.01$) spectroscopic binary are coplanar with their circumbinary disk (misalignment $< 6\degr$ with 68\% confidence), similar to other short-period binary systems. From models of the spectral energy distribution, we found the inner radius of the circumbinary disk ($r_\mathrm{inner} = 0.50 - 0.75\,$au) to be consistent with theoretical predictions of dynamical truncation $r_\mathrm{cav}/a_\mathrm{inner} \approx 3$. The outer orbit of the tertiary star B ($0.40\pm0.28\,M_\odot$, $a\sim65 \pm 18$\,au, $e=0.3\pm0.2$) is not as well constrained as the inner orbit, however, orbits coplanar with the A system are still preferred (misalignment $ < 20\degr$). To better understand the influence of the B orbit on the \tw A circumbinary disk, we performed SPH simulations of the system and found that the outer edge of the gas disk ($r_\mathrm{outer}=8.5\pm0.2\,$au) is most consistent with truncation from a coplanar, circular or moderately eccentric orbit, supporting the preference from the joint orbital fit.
\end{abstract}

\section{Introduction} \label{sec:intro}
The distribution of pre-main sequence multiple system architectures informs our understanding of the mechanisms that govern star and planet formation. Recently, \citet{czekala19} found that the degree of alignment between the disk and its host binary (the mutual inclination, $\theta$) is a strong function of orbital period. Circumbinary disks around short-period binaries ($P < 40$\,days) are preferentially coplanar, while disks around longer period binaries exhibit a wide range of mutual inclinations, including polar configurations ($\theta \approx 90\degr$). These trends might be manifestations of the same physical mechanisms that produce close binaries. Because it is difficult to directly fragment stars on scales $< 5\,\mathrm{au}$ \citep{larson69,bate02,offner16}, it is believed that tight binaries are instead produced from wider binaries that have hardened through star-disk interactions \citep{offner10,bate19}. What mechanisms mediate this evolution, whether migration affects mutual inclination, and whether initial mutual inclination affects migration efficiency are all open questions.

Typically, a nearly coplanar disk surrounding a low eccentricity binary will precess around the binary angular momentum vector as dissipative forces damp the angular momentum vectors of the disk and the binary into alignment \citep{foucart13}. If the binary is sufficiently eccentric, however, then the disk can precess around the eccentricity vector\footnote{The eccentricity vector is drawn from binary apoapse to periapse.} \citep{aly15,martin17,zanazzi18,cuello&giuppone19} and access polar mutual inclinations. Indeed, highly misaligned disks are preferentially found around highly eccentric ($e > 0.7$) binaries \citep{kennedy19,czekala19}.  There are, however, several disks around eccentric \emph{short period} binaries that are coplanar, such as AK~Sco, DQ~Tau, and UZ~Tau E \citep{czekala15a, czekala16, czekala19}. Our understanding of how these period and eccentricity trends interrelate is limited by the sample size of circumbinary disk systems with well-measured architectures.

The pre-main sequence system \tw\ represents an opportunity to expand this sample to aid in the interpretation of binary formation and evolution mechanisms. \tw\ consists of three young \citep[$10\pm3$\,Myr;][]{bell15}, pre-main sequence M3--M4.5 stars in a hierarchical configuration; their spectral types correspond to $\sim 0.3\,M_\odot$ \citep{herczeg14,tofflemire19}. The inner Aa--Ab binary has an orbital period of $P=34.8785\pm0.0009\,$days, an eccentricity of $e=0.628\pm0.006$, and spectral types of M4 and M4.5, respectively \citep{kellogg17}. The Gaia DR2 parallax is $\varpi = 27.31 \pm 0.12$\,mas  \citep[including a $0.02\,$mas systematic term,][]{lindegren18} corresponding to a distance of $36.62 \pm 0.16\,$pc  \citep{gaia18,bailer-jones18}. \citet{tofflemire19} noted that the A and B components suffer from significant excess astrometric noise---possibly due to photometric variability---but the parallax distances for each source are consistent with each other and the A--B orbit arcs in \citet{kellogg17}. Time-series photometry and emission line spectroscopy revealed that accretion from the circumbinary disk to the inner binary is phased with periastron, and that material is preferentially accreted onto the primary star, \tw Aa \citep{tofflemire17,tofflemire19}. The gradual movement of the outer triple companion \citep[projected separation 1\farcs55 or 57\,au;][]{tokovinin15} over a $\sim 20\,$yr baseline suggests an orbital period of $\sim 200 - 800\,$yr \citep{kellogg17}.

\citet{andrews10a} used the Submillimeter Array (SMA) to localize the submillimeter emission in the \tw\ system to the A binary, measuring a flux density of 75\,mJy at 340\,GHz. As demonstrated by fits to the deprojected and azimuthally averaged baselines, the circumbinary disk itself was only marginally resolved (1\farcs11 $\times$ 0\farcs74 beam), but found to have a radius of $\sim20\,\mathrm{au}$. \citet{andrews10a} did not detect \twelve\ $J=3-2$ emission to an upper limit of 0.6\,Jy\,beam${}^{-1}$ integrated over a $0.7\,\kms$ channel. Based upon the fit of an elliptical Gaussian to the visibilities, \citet{andrews10a} derived a disk inclination of $i_\mathrm{disk} = 36\degr \pm 10\degr$ (relative to the sky plane) and disk orientation of $\Omega_\mathrm{disk}=169\degr \pm 15\degr$ (the position angle of the ascending node measured east of north).\footnote{The lack of a gas detection meant degenerate ``flipped'' solutions were also valid: $i_\mathrm{disk} = 144\degr\pm10\degr$ and $\Omega_\mathrm{disk} = 349\degr \pm 15\degr$.}

More recent orbital solutions suggested that the inner binary orbit, circumbinary disk, and outer tertiary orbit may be misaligned. \citet{kellogg17} combined an astrometric observation of the inner binary \citep{anthonioz15} with their double-lined radial velocity solution to constrain the position angle of the ascending node $\Omega_\mathrm{inner} \in [93\degr,123\degr]$ and inclination $i_\mathrm{inner} \in [32\degr,63\degr]$ or $i_\mathrm{inner} \in [118\degr,149\degr]$. These orbital parameters, together with the disk parameters reported in \citet{andrews10a}, suggested that the planes of the spectroscopic binary and the circumbinary disk were misaligned by at least $\imut \sim 30\degr$ \citep{kellogg17}. 

We acquired ALMA observations of the \tw A circumbinary disk to better understand its size and orientation relative the stellar orbits. In \S\ref{sec:data} we describe the ALMA observations and data reduction. In \S\ref{sec:analysis} we dynamically model the gas rotation curve of the \tw A circumbinary disk, fit the spectral energy distribution (SED) of \tw A, and perform a joint stellar orbital fit to the radial velocities and astrometric measurements of the Aa, Ab, and B stars. In \S\ref{sec:discussion} we describe our smoothed-particle hydrodynamics (SPH) simulations of the \tw\ system and discuss how they support our interpretation of the circumbinary disk as nearly coplanar and dynamically truncated both internally and externally by binary companions. We also briefly review similar analog protoplanetary and exoplanetary systems in our discussion of mutual inclination and disk truncation. We conclude in \S\ref{sec:conclusions}.

\section{Data} \label{sec:data}
We obtained deep Atacama Large Millimeter Array (ALMA) observations of the \tw\ system in 2018. We used a correlator setup that assigned two 2 GHz wide spectral windows to the dust continuum (centered on 220\,GHz and 232\,GHz) and two spectral windows at 122\,kHz (0.16\,\kms) resolution to target the \twelve\ and \thirteen\ $J=2-1$ transitions. Two sets of observations (project code 2018.1.01545.S) were executed on Oct 16th and 27th, 2018 (JD 2458408.1404 and 2458419.0483, respectively) using 44 and 43 antennas of the main array, respectively.
The array was similarly configured for each observation, with baselines ranging from 15\,m to 2.4\,km.
Both observing sequences used the J1107-4449 quasar as an amplitude and bandpass calibrator and used the J1126-3828 quasar as a phase calibrator. Each execution spent 47.4 minutes on-source, for a total on-source time of 1\,hr~34.8\,m. The mean precipitable water vapor for each observation was 1.9mm and 0.4mm, respectively.

\begin{deluxetable}{lcc}[t]
\tablecaption{Image properties  \label{tab:ims}}
\tablehead{
\colhead{} &
\colhead{beam dimensions, P.A.} &
\colhead{RMS [mJy beam$^{-1}$]}}
\startdata
226\,GHz cont. & $0\farcs24\times0\farcs19$, -64\degr\ & 0.013 \\
$^{12}$CO $J$=2$-$1 & $0\farcs24\times0\farcs19$, -64\degr\ & 1.1     \\
$^{13}$CO $J$=2$-$1 & $0\farcs25\times0\farcs20$, -63\degr\ & 1.3     \\
\enddata
\tablecomments{The RMS noise levels for the spectral line cubes correspond to the values per 0.8\,\kms\ channel. All images were synthesized with \texttt{robust=0.5}.
}
\end{deluxetable}

\begin{figure*}[t]
\begin{center}
\includegraphics{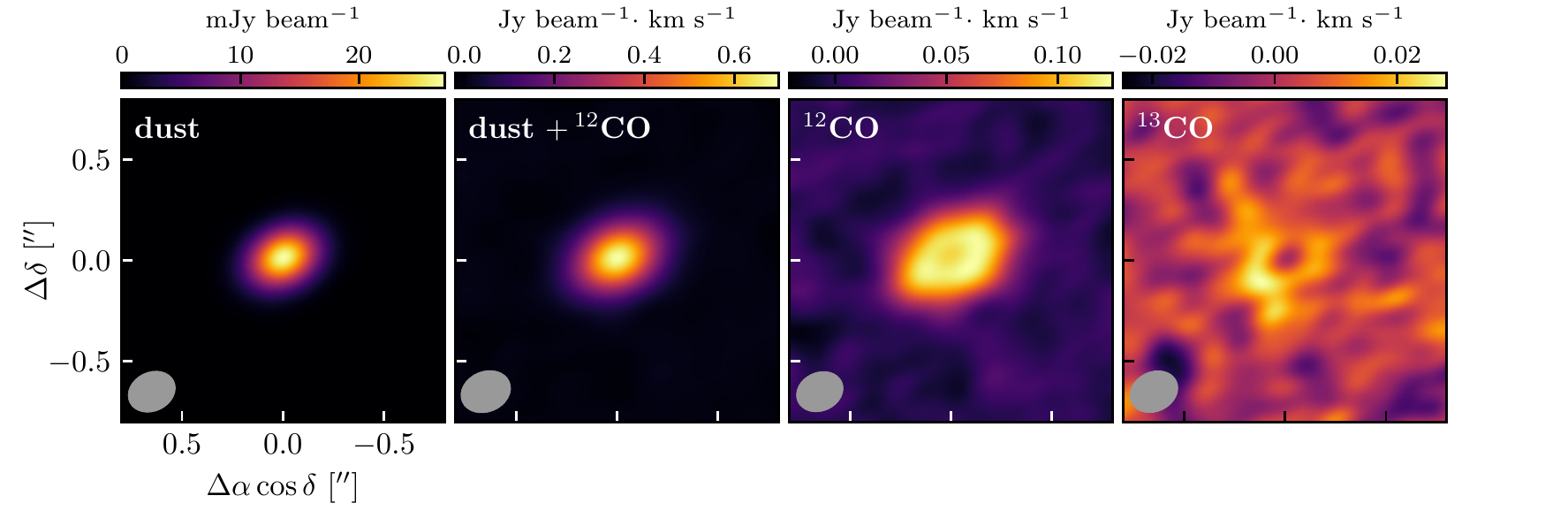}
\figcaption{\emph{from left to right}: the 226\,GHz dust continuum; the channels centered on the \twelve\ emission (including the dust continuum) integrated over the full range of the \twelve\ emission; the continuum-subtracted \twelve\ emission; and the continuum-subtracted \thirteen\ emission.
The FWHM beam is shown in the lower left of each panel. Full channel maps for \twelve\ and \thirteen\ are in \S\ref{subsec:sub-mm-gas} and the Appendix, respectively.
\label{fig:moments}}
\end{center}
\end{figure*}

We began our data reduction with the pipeline-calibrated measurement set provided by ALMA/NAASC staff. We used the CASA 5.4 \citep{mcmullin07} facility software and followed common calibration and imaging procedures \citep[e.g., the DSHARP reduction scripts,][]{andrews18}\footnote{\url{https://bulk.cv.nrao.edu/almadata/lp/DSHARP/}}, the pertinent details of which we now describe. To assess the quality of each execution block we first reduced each observation individually.
We excised the channels with line-emission to create a continuum-only measurement set with a total bandwidth of 4.6\,GHz. We performed an initial round of continuum imaging using the CASA \texttt{tclean} task with \texttt{robust=0.5}, an image size of $512\times512$ pixels, 0\farcs015 pixel size, \texttt{deconvolver="multiscale"}, \texttt{scales=[0, 15, 30, 45, 75]} pixels, a threshold of 0.6\,mJy, and an elliptical mask with position angle 110\degr, semi-major axis of 0\farcs45 and semi-minor axis of 0\farcs36.

We fit the continuum emission with an elliptical Gaussian using the \texttt{imfit} and \texttt{uvmodelfit} tasks. We found excellent astrometric agreement between executions, with the emission centroid located at ICRS 11:10:27.731 -37.31.51.84, coincident with the Gaia position of \tw A to within 0\farcs05. We found adequate agreement between the total continuum fluxes (31.9\,mJy and 36.1\,mJy, respectively), only slightly more different than the expected 10\% amplitude calibration uncertainty.\footnote{See the ALMA Technical Handbook \citep{alma_technical_handbook} and ALMA Memo 594 \url{https://science.nrao.edu/facilities/alma/aboutALMA/Technology/ALMA_Memo_Series/alma594/memo594.pdf}} 

We proceeded to self-calibrate the combined measurement set through a series of applications of \texttt{tclean} to the continuum visibilities using threshold depths \{0.6, 0.15, and 0.15\} $\mathrm{mJy}\,\mathrm{beam}^{-1}$ interleaved with applications of  the \texttt{gaincal} and \texttt{applycal} CASA tasks using spectral-window dependent solves with intervals \{60\,s, 30\,s, and 18\,s\}. We then cleaned to a final depth of $0.02\,\mathrm{mJy}\,\mathrm{beam}^{-1}$ and performed one round of phase and amplitude gain solutions over the scan length (8 minutes, \texttt{solint=`inf'}). We monitored the peak flux, total flux, and RMS of the images throughout the process \citep{brogan18} and found the peak continuum S/N improved to 2130 from an initial value of 260. 

We then used the \texttt{applycal} task to apply the self-calibration solutions to channels containing the spectral line observations\footnote{Though part of a standard self-calibration workflow, we noticed that the S/N of the line channels did not measurably improve after applying the self-calibration solutions. We believe that this is because the fine resolution channels are thermal-noise dominated and not limited by residual phase errors.} (without propagating flags for failed solutions, \texttt{applymode=calonly}). We estimated the continuum from nearby line-free channels and subtracted it from the spectral line observations using the \texttt{uvcontsub} CASA task. The line channels were imaged using the \texttt{tclean} task with the \texttt{auto-multithresh} masking algorithm with \texttt{cell="0.015arcsec"}, \texttt{gain=0.1}, \texttt{deconvolver="multiscale"}, \texttt{scales=[0, 10, 30, 100, 200, 300]} pixels, \texttt{robust=0.5}, and deconvolved to a depth of \texttt{threshold="0.1mJy"}. The beam dimensions (FWHM) and image-plane RMS are described in Table~\ref{tab:ims}.

\begin{figure}[t]
\begin{center}
\includegraphics{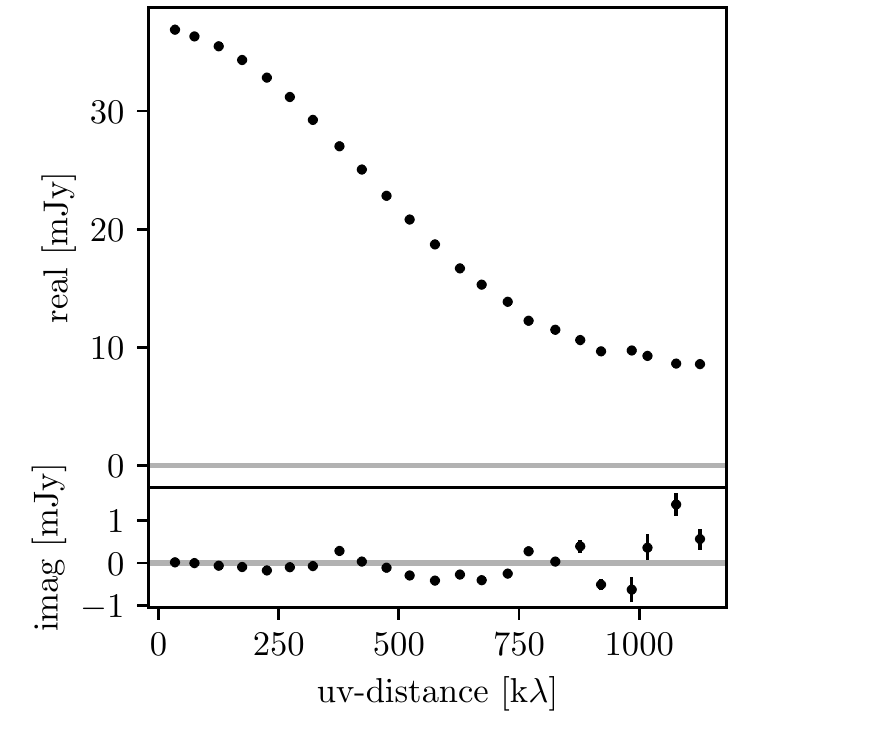}
\figcaption{The continuum visibilities deprojected from the disk inclination. That the flux drops with increasing uv distance indicates that the continuum emission is spatially resolved. The Gaussian profile with $\mathrm{FWHM}\approx 1200\,\mathrm{k}\lambda$ implies an image plane morphology with $\mathrm{FWHM} \approx 0\farcs15$, or 5.5\,au. 
\label{fig:deprojected}}
\end{center}
\end{figure}

We summed the pixels within the continuum CLEAN mask to measure a total continuum flux of 36.7\,mJy. The dust continuum emission is compact, with nearly all of the flux contained within the central beam. We fit elliptical Gaussians to the continuum emission with the \texttt{uvmodelfit} and \texttt{imfit} tasks and derived FWHM dimensions of (0\farcs12 $\times$ 0\farcs12) and (0\farcs16 $\times$ 0\farcs10), respectively. Since these dimensions are on the order of the beam size, to better quantify whether we resolved the dust continuum we also deprojected and azimuthally averaged the continuum visibilities using the \texttt{uvplot} package \citep{uvplot_mtazzari}, shown in Figure~\ref{fig:deprojected}. We used deprojection values of $i_\mathrm{disk}=49\degr$ and $\Omega_\mathrm{disk}=116.5\degr$, which were derived from our modeling effort of the \twelve\ $J=2-1$ line emission, since the more extended emission provided better constraints for these parameters than the dust continuum emission (see \S\ref{subsec:sub-mm-gas}). The declining visibility profile demonstrates that the outer extent of the disk is resolved: if the deprojected continuum emission profile is represented by a Gaussian, then a Fourier-domain  $\mathrm{FWHM}\approx 1200\,\mathrm{k}\lambda$ implies that the image plane Gaussian has $\mathrm{FWHM} \approx 0\farcs15$, or 5.5\,au. 

In preliminary line imaging, we identified \twelve\ emission from [-9, 10]\,\kms (LSRK). To save computational complexity, we used the task \texttt{mstransform} to average the channels to 0.8\,\kms\ width. The \twelve\ emission was strongly detected across this velocity range (peak channel S/N$=30$, see Figure~\ref{fig:chmaps}, top panel); the \thirteen\ emission was detected at lower but still significant levels (peak channel S/N$=6$, see Appendix~\ref{sec:appendix}). We used the \texttt{immoments} CASA task to sum all flux in each channel across the velocity dimension, producing the moment maps in the third and fourth panels of Figure~\ref{fig:moments}. We used the \texttt{imstat} CASA task to sum all flux within the CLEAN mask across the velocity and and spatial dimensions, yielding integrated fluxes of 526\,mJy\,\kms\ and 86\,mJy\,\kms\ for \twelve\ and \thirteen, respectively.

We noticed a central cavity in the moment maps of the continuum-subtracted line emission, corresponding to the location of peak dust continuum. The depression is most apparent in the \twelve\ emission but is also visible in the \thirteen\ emission (Figure~\ref{fig:moments}, third and fourth panels). To investigate whether this may be a continuum subtraction artefact, we produced channel maps and a moment map for the non-continuum subtracted \twelve\ spectral channels (Figure~\ref{fig:moments}, second panel). This moment map does not exhibit a central cavity, suggesting that the feature seen in \twelve\ and \thirteen\ is indeed a continuum subtraction artefact. In protoplanetary disks, such an artefact can arise when gas emission on the near-side of the disk is optically thick and absorbs continuum emission originating from the disk midplane. When the continuum emission (estimated from channels offset in velocity from the line emission) is subtracted, most if not all of the line flux is also (erroneously) subtracted \citep[for a full description of the effect, see][]{weaver18}. For this artefact to be significant, the continuum emission needs to have a brightness temperature comparable to the line emission, suggesting that the continuum emission is also optically thick, or nearly so.

No noticeable continuum or gas emission is present in the system beyond the disk surrounding \tw A.
We used an aperture approximately three times the area of the beam to extract continuum photometry at the location of B \citep[see \S\ref{subsec:stellar-orbits};][]{mason18}, and did not detect anything ($37 \pm 27$\,$\mu$Jy).

\section{Analysis}
\label{sec:analysis}

\subsection{Dynamical gas analysis}
\label{subsec:sub-mm-gas}

Following the approach in \citet{czekala15a}, we constructed a forward model of the continuum-subtracted \twelve\ $J=2-1$ visibilities to derive constraints on the disk architecture and velocity field. Briefly, a 3D model of the disk density, temperature, and velocity is parametrically defined and then ray-traced using \texttt{RADMC-3D} \citep{dullemond12} to produce image cubes. These cubes are Fourier-transformed and sampled at the spatial frequencies corresponding to the baselines of the ALMA measurement set to compute the likelihood of the dataset.\footnote{Using the \texttt{DiskJockey.jl} package \citep{czekala15a}, \url{https://github.com/iancze/DiskJockey}} The posterior of the model parameters, defined by the data likelihood and any additional prior probability distributions, is explored using Markov Chain Monte Carlo (MCMC).

\begin{figure*}[t]
\begin{center}
\includegraphics{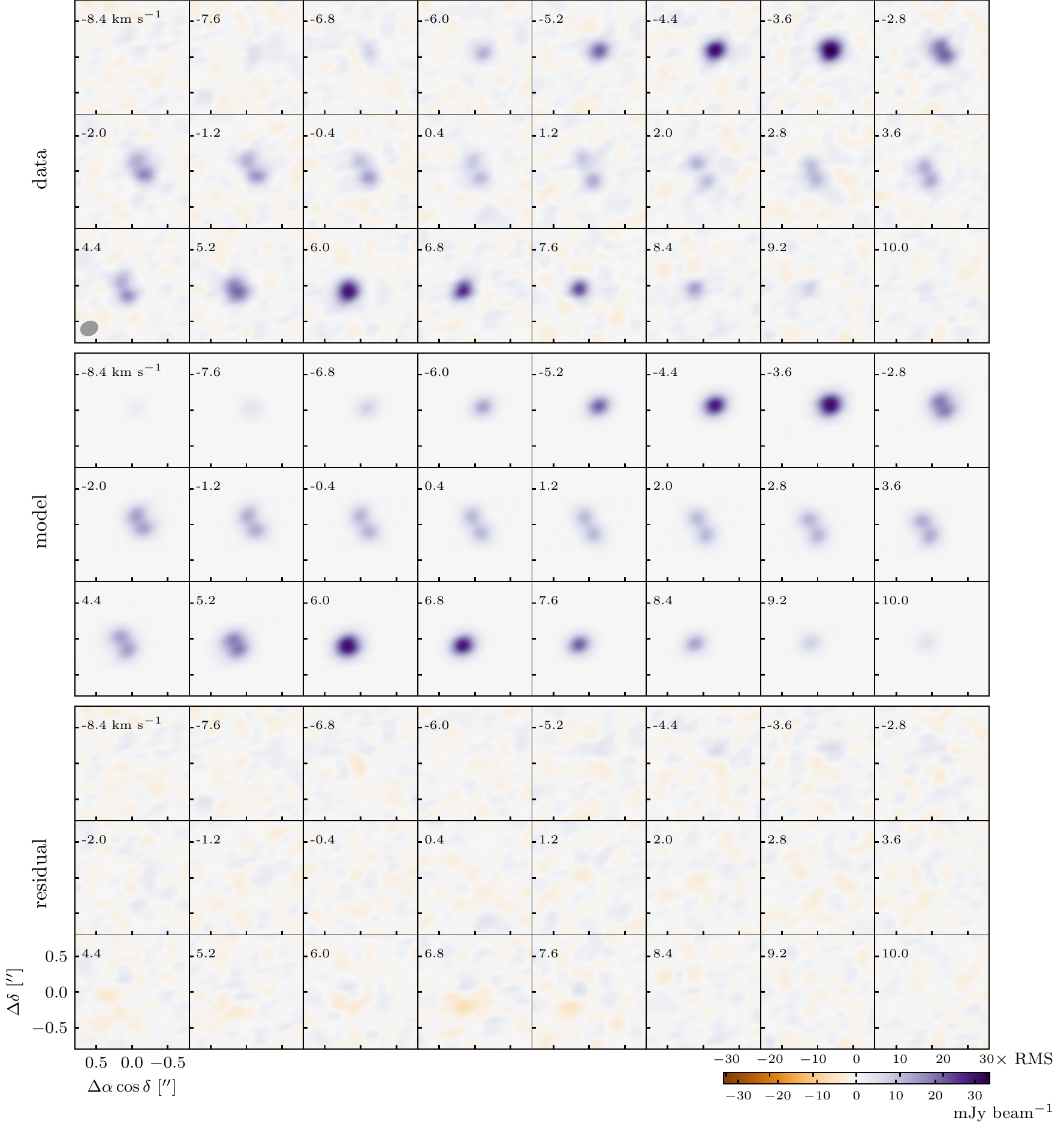}
\figcaption{Continuum-subtracted \twelve\ $J=2-1$ data, model, and residual channel maps. Model and residual visibilities were imaged the same way as the data. Velocity scale is labeled in the LSRK frame.
\label{fig:chmaps}}
\end{center}
\end{figure*}

We initially explored models with self-similar surface density profiles $\Sigma(r)$ of the form described in \citet{czekala15a} and references therein \citep[see also][]{sheehan19}. We found that the gradual exponential taper of this profile at large radii was not well-matched to the \tw A \twelve\ $J=2-1$ emission, which appears to decay quickly at large radii (see Figure~\ref{fig:moments}). Drawing inspiration from dust continuum studies which flexibly parameterized the surface brightness profile $I(r)$ using the Nuker function \citep[e.g.,][]{tripathi17}, we adapted this to a surface density profile as
\begin{equation}
\Sigma(r) = \Sigma_c \left(\frac{r}{r_c} \right)^{-\gamma} \left[1 + \left ( \frac{r}{r_c} \right)^\alpha \right]^{(\gamma - \beta)/\alpha}.
\label{eqn:nuker}
\end{equation}
We followed \citet{tripathi17} and chose to sample in $\log_{10}\alpha$ and imposed priors on shape parameters $\log_{10}\alpha$, $\beta$, and $\gamma$. We restricted $\log_{10}\alpha$ and $\beta$ with uniform probability to the ranges of $0 \leq \log_{10} \alpha \leq 2$ and $2 \leq \beta \leq 10$, respectively, and imposed a tapered prior on $\gamma$ of the form 
\begin{equation}
    p(\gamma) = \frac{1}{1 + e^{-5 (\gamma + 3)}} - \frac{1}{1 + e^{-15 (\gamma - 2)}}.
\end{equation}
As discussed in \citet{tripathi17}, these priors are practically motivated to allow a broad range of surface density profiles, including those with interior cavities and sharp outer edges, while restricting the parameter space sufficiently to avoid pathological surface density profiles ill-suited to protoplanetary disks. Because the Nuker profile increases the dimensionality of the posterior by adding two new parameters, we kept computational demands tractable by zeroing out the phase-center offsets $\delta_\alpha$, $\delta_\delta$ typically used with the standard model.  

We sampled the posterior distribution using the \texttt{DiskJockey} package \citep{czekala15a} and assessed convergence both visually and by applying the Gelman-Rubin convergence diagnostic to independent chain ensembles. The resulting marginal posteriors on the disk model parameters are listed in Table~\ref{table:components}. Most marginal distributions are well-described by Gaussians. The posteriors on the temperature profile exponent $q$ and Nuker shape parameters $\log_{10} \alpha$ and $\beta$ ran up against the range of their prior bounds, so $68\%$ upper or lower confidence intervals are quoted for these parameters instead. Figure~\ref{fig:chmaps} shows a realization of the model and residual visibilities drawn from the posterior distribution, imaged in the same way as the data. The residual channel maps are broadly consistent with residual thermal noise, demonstrating that the synthesized model is a good fit to the data. We define the outer edge of the disk using the radius that contains 95\% of the mass: $r_\mathrm{outer} = 8.5 \pm 0.2$ au. That the 226\,GHz continuum emission (Figure~\ref{fig:moments}) is more compact than the extent of the gas emission is consistent with the expectation that radial drift has moved large dust grains (mm or cm sized) inward \citep{andrews20}.

\begin{deluxetable}{lc}
\tablecaption{Inferred Disk Parameters \label{table:components}}
\tablehead{\colhead{Parameter} & \colhead{Value}}
\startdata
$M_\mathrm{A}\quad [M_\odot]$ & $0.534 \pm 0.010$ \\
$r_c$ [au] & $6.8 \pm 0.2$\\
$T_{10}$ [K] & $38 \pm 1$ \\
$\gamma$ & $-3.5 \pm 0.5$ \\
$q$ & $\leq 0.05$ \\
$\log_{10} \alpha$ & $\geq 1.6$ \\ 
$\beta$ & $\geq 9.8$\\
$\log_{10} M_\mathrm{disk} \quad \log_{10} [M_\odot]$ & $-6.37 \pm 0.09$ \\ 
$\xi\,[\kms]$ & $0.60 \pm 0.03$\\ 
$i_\mathrm{disk} \quad$ [\degr] & $48.8 \pm 0.7$\tablenotemark{a}\\ 
$\Omega_\mathrm{disk}$ [\degr] & $116.5 \pm 0.4$\\ 
$v_r$\,$[\kms]$ & $1.22\pm 0.02$\tablenotemark{b}\\ 
\enddata 
\tablenotetext{a}{Ambiguity with $i_\mathrm{disk}=131.2\pm0.7\degr$}
\tablenotetext{b}{LSRK reference frame.}
\tablecomments{Using \twelve.
	The 1D marginal posteriors are well-described by a Gaussian, so we report symmetric error bars here (statistical uncertainties only).
	These parameters were inferred using a distance of $d = 36.62\,$pc.
}
\end{deluxetable} 

The inferred surface density profile constrains the peak density to approximately 7\,au in radius; the $q\approx 0$ temperature exponent indicates that the gas in the ring is nearly constant temperature across its narrow radial extent. However, the gas-depleted central cavity implied by this surface density profile is not real, but instead reflects the continuum-subtraction artefact that removed \twelve\ $J=2-1$ emission from central radii. As shown in Figure~\ref{fig:chmaps}, the flexible Nuker profile provided an excellent fit to the continuum-subtracted \twelve\ $J=2-1$ emission and allowed us to achieve our primary objectives of inferring the disk velocity field and measuring the outer extent of the circumbinary disk. A joint dust and gas model simultaneously fit to the dust continuum and \twelve\ $J=2-1$ emission could in principle recover a more accurate gas surface density profile. However, the behavior of the surface density profile at small radii is effectively a nuisance component to our dynamical analysis and does not justify a more sophisticated model, especially since its significantly expanded computational requirements would curtail our ability to thoroughly explore the posterior distributions of key parameters ($M_\mathrm{A}$, $i_\mathrm{disk}$, and $\Omega_\mathrm{disk}$) via MCMC.

We constrained the disk position angle to be $\Omega_\mathrm{disk} = 116.5\degr \pm 0.4\degr$, which is significantly different from the value found by \citet{andrews10a} ($\Omega_\mathrm{disk} = 144\degr \pm 10\degr$). We constrained the disk inclination to be either $i_\mathrm{disk} = 48.8\degr \pm 0.7\degr$ or $i_\mathrm{disk} = 131.2\degr \pm 0.7\degr$, which is also in disagreement with the values found by \citet{andrews10a} ($i_\mathrm{disk} = 36\degr \pm 10\degr$ or $i_\mathrm{disk} = 144\degr \pm 10\degr$). \citet{andrews10a} derived the disk inclination and position angle by fitting an elliptical Gaussian to marginally resolved sub-mm continuum observations, so it is only mildly surprising that their simplistic model deviates from the new values we derived using a more realistic dynamical gas model fit to higher quality observations. We constrained the central stellar mass to be $M_\mathrm{A} = M_\mathrm{Aa} + M_\mathrm{Ab} = 0.534 \pm 0.010\,M_\odot$ and the systemic velocity of the \tw A circumbinary disk to be $1.22\pm 0.02\,\kms$ in the LSRK frame ($10.48\pm 0.02\,\kms$ in the BARY frame).\footnote{In the direction of TWA~3, $v_\mathrm{LSRK} = v_\mathrm{BARY}- 9.26\,\kms$.} 
This is fully consistent with the radial velocity of the \tw A barycenter $v_\mathrm{LSRK} = 0.91\pm0.40\,\kms$ \citep[$\gamma_\mathrm{A} = 10.17\pm0.40\,\kms$ on the CfA system,][]{kellogg17}.

\subsection{SED modeling and (sub)mm spectral index} \label{subsec:sed}
\begin{figure}[t]
\begin{center}
\includegraphics[width=3.5in]{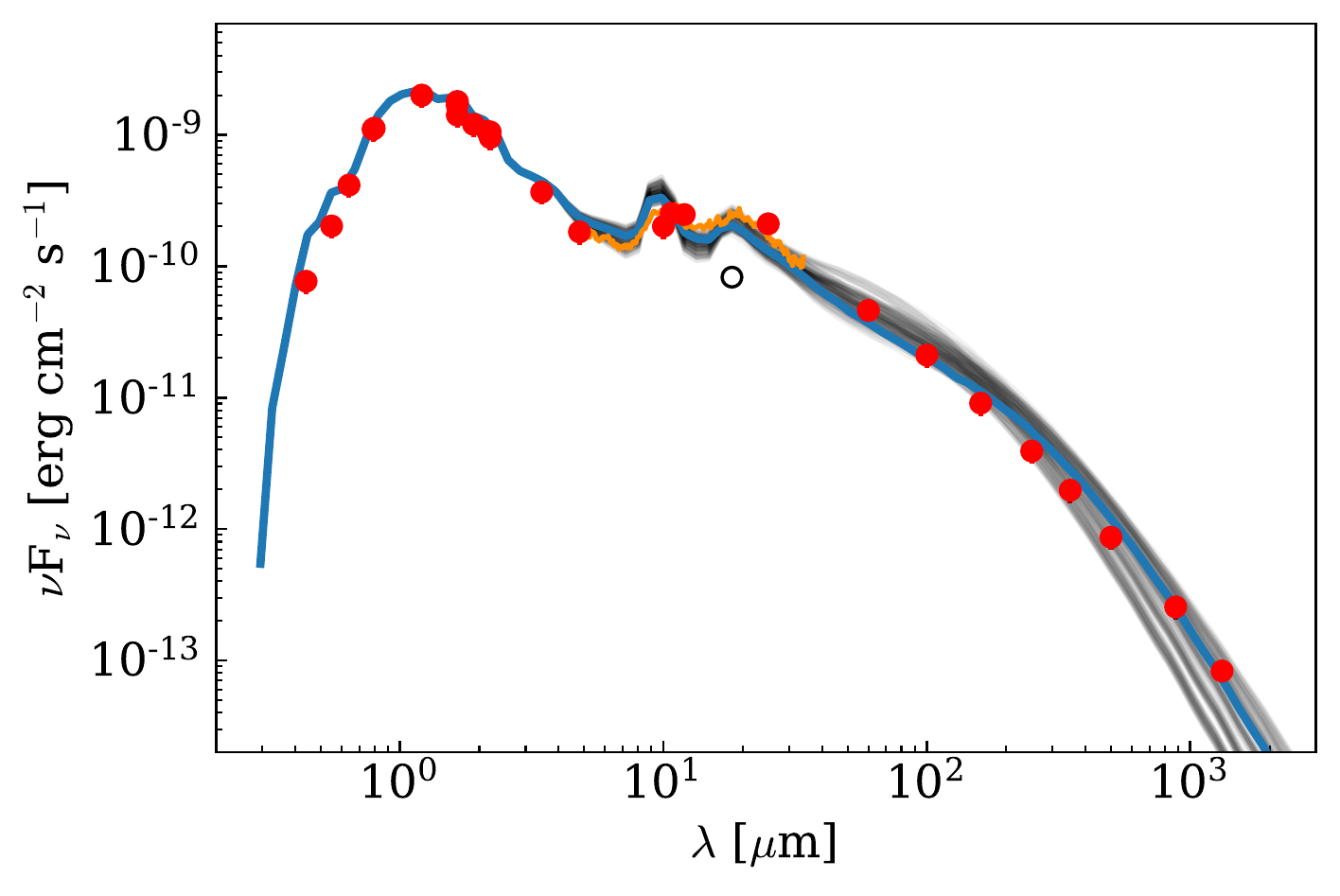}
\figcaption{The SED of \tw A: Photometric observations are shown in red, and the \emph{Spitzer}/IRS spectrum in orange. The Q-band observation (black empty circle) from \citet{jayawardhana99} was not used during the fitting. The best-fit model is shown in blue along with the 100 highest likelihood SED models (grey).
\label{fig:SED_models}}
\end{center}
\end{figure}

In light of our new 226\,GHz ALMA flux density measurement, we updated the spectral energy distribution (SED) of the \tw A system to learn about the properties of the circumbinary disk and its interior cavity. We sourced photometric fluxes from the SED compilation in \citet{kellogg17}. We also incorporated the mid-IR spectrum from the IRS spectrograph onboard the \emph{Spitzer Space Telescope} \citep{andrews10a} with the contribution from the B component subtracted. The SED (shown in Figure~\ref{fig:SED_models}) continuously decreases redward of $\sim$20\,$\mu$m, suggesting that the circumbinary disk is truncated at larger radii, highly settled, or both. The observed spectral slope between the SMA 880\,$\mu$m and the 1.3\,mm ALMA data is:
\begin{equation}
\alpha = \frac{\mathrm{d}\log F_\nu}{\mathrm{d} \log \nu} = 1.7\pm 0.3,
\end{equation} 
where the uncertainty is calculated by adopting a 10\,\% calibration uncertainty for both points. That the spectral index is $\alpha \lesssim 2$ suggests the dust continuum emission is optically thick at these wavelengths, consistent with the continuum subtraction artifact described in \S\ref{sec:data}. If the maximum grain size in the disk is between 500$\,\mu$m and 1\,cm, self-scattering from high albedo dust grains can reduce the (sub)mm-wave emission from an optically thick region and produce $\alpha < 2$ \citep{zhu19}.

Motivated by the near-infrared dip ($\lambda \sim 10\,\mu$m) in the SED, we followed \citet{andrews10a} and constructed a simple disk model with two zones: an inner cavity with a constant surface density from $r_\mathrm{in}$ to $r_\mathrm{cav}$, and a disk with surface density profile $\Sigma(r) \propto 1/r$ from $r_\mathrm{cav}$ to $r_\mathrm{out}$. We fixed $r_\mathrm{in}$ to 0.2\,au, following \citet{kellogg17}'s constraint on the inner binary semi-major axis of $0.19\,$au. We fixed $r_\mathrm{out}= 8.5\,$au and $i_\mathrm{disk} = 49\degr$ following our analysis of the \twelve\ $J=2-1$ line emission. Our model had five free parameters: total disk dust mass $M_\mathrm{dust}$, cavity radius $r_\mathrm{cav}$, cavity depletion factor $\delta$, flaring parameter $\beta$, and scale height at 10 au $H_{10}$. The surface density within the cavity ($r_\mathrm{in} < r < r_\mathrm{cav}$) was set to a constant surface density $\Sigma_{\rm cav} = \delta\,\Sigma_\mathrm{disk}(r_\mathrm{cav})$. The scale height as a function of radius is defined as 
\begin{equation}
    H(r) = H_{10} \left ( \frac{r}{10\,\mathrm{au}} \right )^\beta.
\end{equation}

We followed the approach in \citet{andrews10a} and combined the Aa and Ab components into a single stellar photosphere with an effective temperature of 3350\,K and radius of 1\,R$_\odot$ (updated to the new \emph{Gaia} distance), equivalent to 0.11\,L$_\odot$. We adopted the dust composition values used in \citet{pinte2016}: the grain size distribution followed $\mathrm{d}n/\mathrm{d}a \propto a^{-3.5}$, where $a$ is the grain size. The distribution ranged from $a_\mathrm{min}=0.01\,\mu$m to $a_\mathrm{max}$=1\,cm. We computed model SEDs using the MCFOST radiative transfer code \citep{mcfost} assuming no interstellar extinction \citep{mcjunkin2014}. Our grid of models spanned parameter ranges $M_{\rm dust}$: [0.5, 1, 2, 4, 8, 16, 32, 64, 128] $\times 10^{-6}\, M_\odot$; $r_\mathrm{cav}$: [0.2, 0.5, 0.75, 1, 1.5, 2]\,au; $\delta$: 1/[1, 10, 100, 1000]; $\beta$ : [1.025, 1.05, 1.075]; and $H_{10}$: [0.2, 0.3, 0.4, 0.5, 0.6]\,au. 

For each model, we calculated a $\chi^2$ figure of merit using the observed photometry and IRS spectrum. Since we were only interested in disk properties, we only used photometric points $\lambda > 1\,\mu \mathrm{m}$ in the following SED fit. We also excluded the Q-band (18.2\,$\mu$m) observation from \citet{jayawardhana99} from the fit since it is an outlier compared to the other photometric points. A set of consistently calibrated photometric and spectroscopic covariance matrices does not exist for the \tw A spectroscopic dataset, and so we were unable to use per-datapoint flux uncertainties in our construction of a $\chi^2$ fit metric. Instead, we explored the consistency of the model grid with the SED data by the following procedure.

First, because adjacent pixels in spectroscopic fluxes are frequently correlated due to residual calibration errors, we subsampled the spectrum and only fit every third point. Then, we explored the consistency of the model grid with the SED data by assigning relative uncertainties of 5\%, 10\%, and 20\% for each photometric and spectroscopic datapoint and calculated the $\chi^2$ metric. We found that $r_\mathrm{cav}=0.20$\,au was excluded at high significance ($>$ 99\% probability) for all choices of uncertainty reweighting factors. The 20\% reweighting for both photometric and spectroscopic datasets yielded the lowest reduced $\chi_\nu^2$ of 1.16. Based on the models with high figures of merit (see Figure~\ref{fig:SED_models}) we conclude that the disk around TWA~3A has a dust mass of $1-8 \times 10^{-6}M_\odot$ and a disk cavity $r_{\rm cav}$ of $\approx$0.5-0.75\,au. The other model parameters were not well constrained. 

\subsection{Stellar orbits}
\label{subsec:stellar-orbits}
From the literature, we collected a diverse orbital dataset including radial velocity and astrometric measurements of all three stars in the \tw\ hierarchical triple \citep{reipurth93,webb99,weintraub00,brandeker03,correia06,janson14,tokovinin15,anthonioz15,kellogg17,knapp18,mason18}.\footnote{Following \citet{kellogg17}, we assigned a date of 1992.0216 to the observation by \citet{reipurth93}.} Our goal was to extend the comprehensive analysis of \citet{kellogg17} by incorporating the new disk-based dynamical constraint on $M_A$ (see \S\ref{subsec:sub-mm-gas}) into a joint hierarchical triple fit with the extant radial velocity and astrometric datasets and \emph{Gaia} parallax. 

We modelled these diverse datasets using the \texttt{exoplanet} software package \citep{exoplanet}. Briefly, \texttt{exoplanet} is designed to unify routines needed for orbital parameter inference within the \texttt{PyMC3} \citep{pymc3} framework. Posterior gradients are provided through the \texttt{Theano} framework \citep{theano}, enabling the usage of powerful MCMC samplers like Hamiltonian Monte Carlo \citep[HMC;][]{hoffman11} to efficiently explore high dimensional spaces. To fit the \tw\ datasets, we extended \texttt{exoplanet} to include functionality for astrometric orbits; these routines have been available in the main \texttt{exoplanet} package as of v0.2.0. We constructed the hierarchical model by nesting a Keplerian orbit for Aa--Ab (``inner'') inside of a wider orbit for A--B (``outer''). In the following analysis, we adopted orbital conventions where the argument of periastron $\omega$ is reported as the value of the ``primary'' star and $\Omega$ describes the position angle of the ascending node, which is the node where the secondary is \emph{receding} from the observer. For the inner orbit, $\omega_\mathrm{Aa}$ refers to the argument of periastron of \tw Aa and $\Omega_\mathrm{inner}$ refers to the position angle of the \tw Aa--Ab ascending node. For the outer orbit, $\omega_\mathrm{A}$ refers to the argument of periastron of \tw A (under the assumption that the Aa and Ab stars can be treated as a single star, A) and $\Omega_\mathrm{outer}$ refers to the position angle of the \tw A--B ascending node.

Following standard radial velocity analysis, we included ``jitter'' and offset terms for each of the instruments. In keeping with \citet{kellogg17}, we derived orbital parameters using the CfA radial velocity reference scale. Because there may be a small but unknown systematic radial velocity offset between the CfA and ALMA velocity scales, we did not use the systemic velocity of the \tw A circumbinary disk (at the epoch of the ALMA measurement) in the joint model. We applied uniform priors on the following quantities: $\log P_\mathrm{inner}$, $\log P_\mathrm{outer}$, $\cos i_\mathrm{inner}$, $\cos i_\mathrm{outer}$, and $\log$ jitter
terms. We applied broad Gaussian priors on the sampled stellar masses of $M_\mathrm{Ab}$ of $0.29 \pm 0.50\,M_\odot$ and $M_\mathrm{B}$ of $0.3 \pm 0.5\,M_\odot$, loosely corresponding to the spectral types, and truncated to positive values only. We also applied broad Gaussian priors of $0.0 \pm 5.0\,\kms$ on the instrument offset terms.

\input{table.tex}

\begin{figure}[t]
\begin{center}
\includegraphics{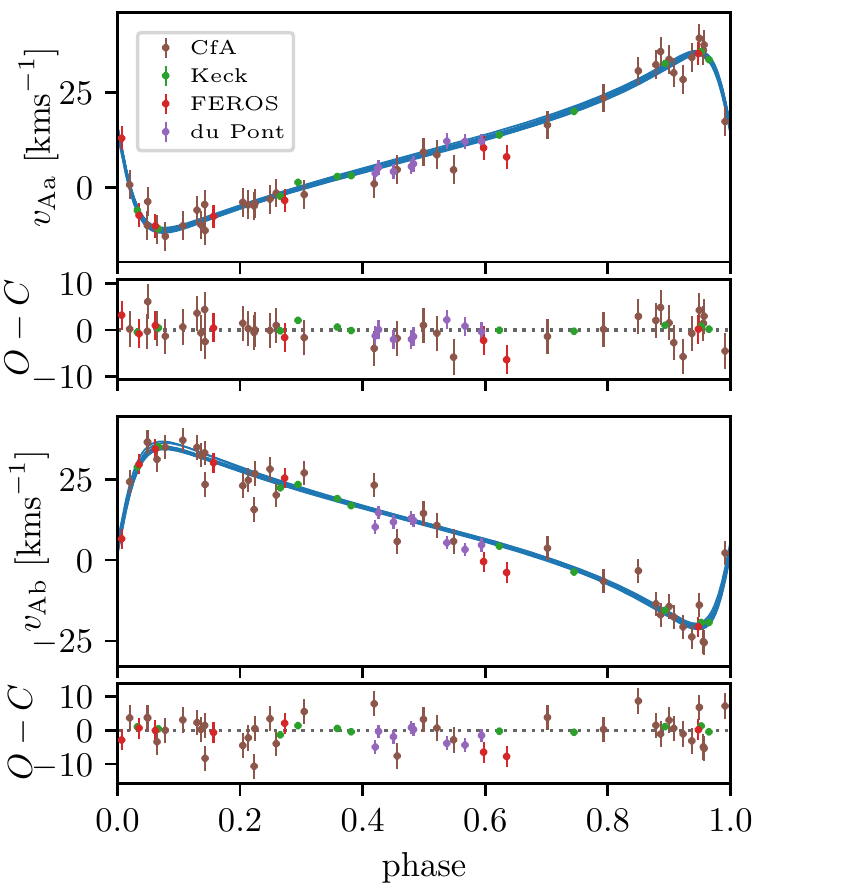}
\figcaption{The phase-folded inner binary orbit and radial velocity residuals. In blue are ten realizations of the inner orbit with the velocity trend from B removed. The small scatter demonstrates that the inner orbital parameters are tightly constrained by the data.
\label{fig:sb_orbit}}
\end{center}
\end{figure}

Table~\ref{tab:orbits} lists a full description of the inferred orbital parameters, where posterior means and standard deviations are provided for parameters whose posteriors are approximately Gaussian. Table~\ref{tab:orbits} covers two scenarios: 1) the ``primary'' solution where $i_\mathrm{inner} > 90\degr$ and 2) the ``alternate'' solution where $i_\mathrm{inner} < 90\degr$. The alternate solution yields retrograde orbits between the inner and outer orbits. We show the phase-folded spectroscopic binary orbit (identical for both scenarios) in Figure~\ref{fig:sb_orbit}, which is in good agreement with that found by \citet{kellogg17}. The joint hierarchical fit delivered more precise posteriors for many outer orbital parameters, though some of the degeneracies noted in \citet{kellogg17} still remain.

\begin{figure}[t]
\begin{center}
\includegraphics{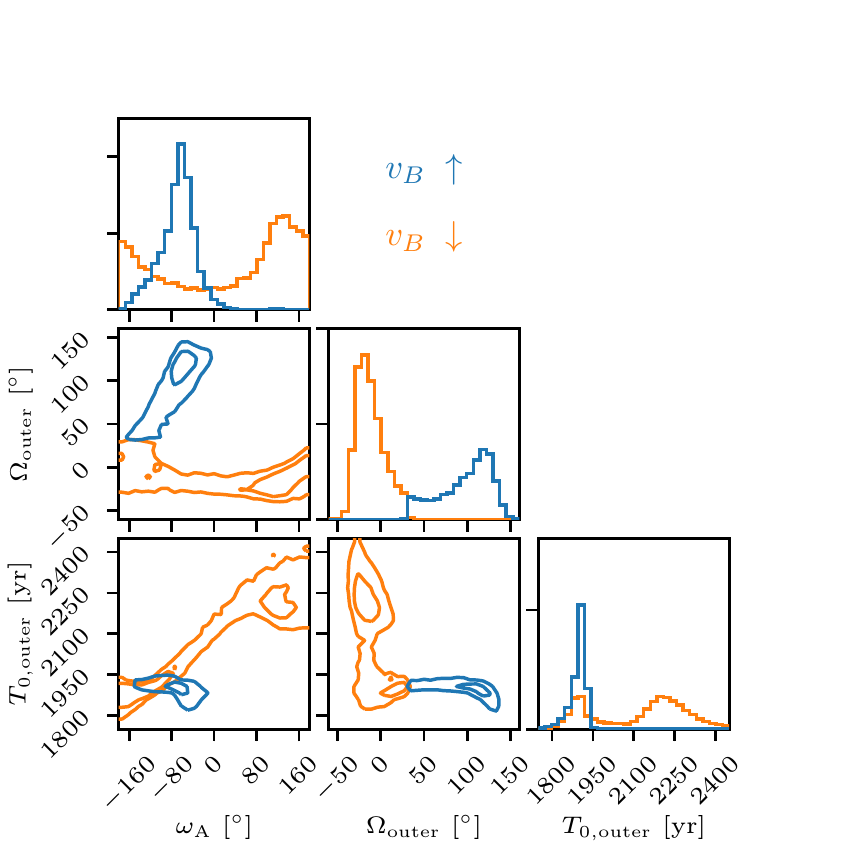}
\figcaption{Corner plot of the sampled outer orbital parameters that have bimodal posterior distributions; contours are 1$\sigma$ and 2$\sigma$ within each mode. This figure is identical for both the ``primary'' and ``alternate'' solutions. Samples are color-coded based on whether they deliver increasing (blue) or decreasing (orange) $v_\mathrm{B}$ velocities over the Keck measurement baseline (see Figure~\ref{fig:astrometry}). Note that the full posterior (the sum of the blue and orange contours) was sampled simultaneously, the samples have been bifurcated for plotting purposes only.
\label{fig:corner}}
\end{center}
\end{figure}

\begin{figure*}[t]
\begin{center}
\includegraphics[width=\linewidth]{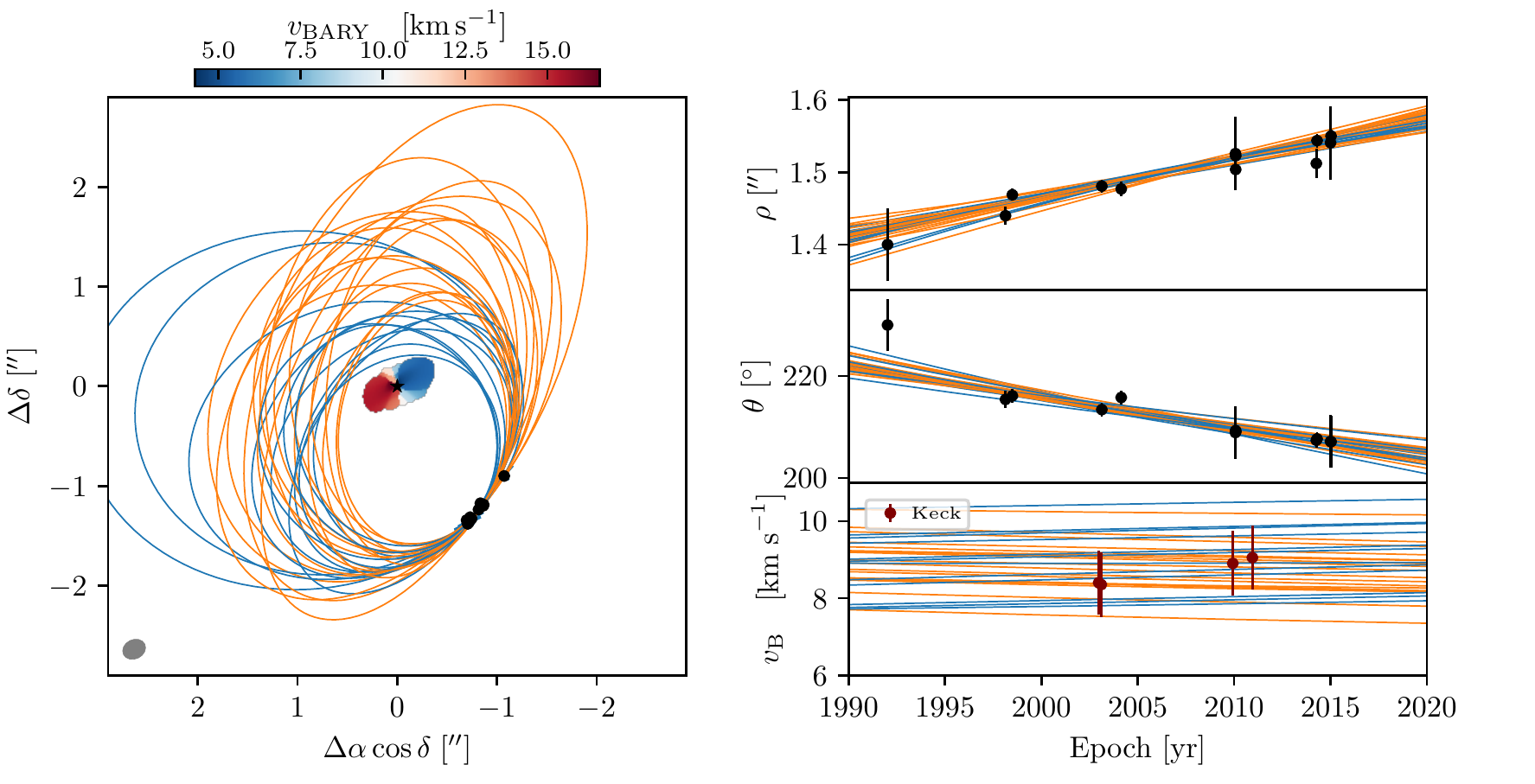}
\figcaption{Thirty representative outer orbits drawn from the highlighted posterior modes in Figure~\ref{fig:corner}.  \emph{left}: the sky plane, centered on \tw A (Aa and Ab are represented by the black star), with the the velocity field of the surrounding circumbinary disk. \emph{right}: the astrometric data and \citet{kellogg17} Keck measurements of $v_B$. Orbits in blue correspond to solutions that deliver increasing $v_B$ over the Keck observation baseline; orange orbits are decreasing.
\label{fig:astrometry}}
\end{center}
\end{figure*}

Three outer orbit parameters have bimodal posterior distributions: $\omega_\mathrm{A}$, $\Omega_\mathrm{outer}$, and $T_{0,\mathrm{outer}}$. The posteriors for these outer orbit parameters are identical between the primary and alternate solutions, so the single corner plot in Figure~\ref{fig:corner} is valid for both scenarios. Representative outer orbits drawn from the posterior distribution are shown in Figure~\ref{fig:astrometry}. Although the formal uncertainties and inferred jitter values of the Keck $v_\mathrm{B}$ measurements are large ($0.59\,\kms$ and $0.82\,\kms$, respectively), the actual scatter of the four measured values is substantially smaller. The first two measurements (separated by 48 days in 2002/2003) differ by only $0.05\,\kms$. The second two measurements (separated by one year in 2009/2010) differ by only $0.15\,\kms$. While circumstantial, this does raise the possibility that the uncertainties on the $v_\mathrm{B}$ measurements are overestimated (potentially driven by the scatter in $v_\mathrm{Aa}$ and $v_\mathrm{Ab}$) and that the increase in $v_\mathrm{B}$ over the Keck measurement baseline ($\approx 0.5\,\kms$ over 8 years) may be significant. If true, a monotonically increasing $v_\mathrm{B}$ clearly favors a single posterior mode, highlighted in blue in Figures~\ref{fig:corner}~\&~\ref{fig:astrometry} (the mode corresponding to decreasing $v_\mathrm{B}$ is highlighted in orange). The MCMC samples and PyMC3 models corresponding to both scenarios are available online.\footnote{\url{https://zenodo.org/record/4568830\#.YDvTb11Kida} \citet{samples} and \url{https://github.com/iancze/TWA-3-orbit}}

We used these orbital posteriors and the inferred values of $i_\mathrm{disk}$ and $\Omega_\mathrm{disk}$ to calculate the mutual inclinations $\theta$ between the disk and the stellar orbits \citep[e.g., Equation 1,][]{czekala19} under different combinations of the degenerate orientations.\footnote{Technically there are yet two more degenerate scenarios where $i_\mathrm{disk} < 90\degr$ but $i_\mathrm{inner} > 90\degr$, or vice-versa. Since the inferred inclinations of the inner binary and circumbinary disk are already so similar (modulo the degeneracy), we do not consider these scenarios.} As discussed in \citet{czekala19}, the sensible imposition of spherically isotropic priors (i.e., the uniform priors on $\cos i_\mathrm{disk}$ and $\cos i_\mathrm{inner}$) results in effective mutual inclination priors of $p(\theta) \propto \sin(\theta)$. These isotropic priors have the consequence of strongly disfavoring coplanar architectures---simply because of the small phase space volume. To quantify the constraining power of the data only, we also report the posteriors re-weighted such that the effective prior is flat (i.e., similar to the marginal likelihood  $p(\mathrm{data} |\,\theta$)). 

\input{mut.tex}

\section{Discussion} \label{sec:discussion}

\subsection{Mutual inclinations}
As established in \S\ref{subsec:stellar-orbits}, the inner binary Aa--Ab orbit and circumbinary disk are very nearly coplanar with each other: across all scenarios listed in Table~\ref{tab:mut}, $\theta_\mathrm{disk-inner} \lesssim 6\degr$ under a flat mutual inclination prior ($\theta_\mathrm{disk-inner} \lesssim 13\degr$ under $\sin \theta_\mathrm{disk-inner}$ prior). \citet{czekala19} found that circumbinary planets, debris disks, and protoplanetary disks around short-period binaries ($P \lesssim 40\,\mathrm{days}$) all have low mutual inclinations. So, the low mutual inclination for the \tw A is consistent with expectations given its 35-day orbital period. 

That circumbinary disk mutual inclination trends with binary period is likely a byproduct of the formation mechanism for tight binary stars, which requires formation at larger distances and migration to present-day configurations \citep{bate02}. The close binary fractions of T Tauri stars and field stars are similar \citep{kounkel19}, implying that this migration occurs quickly, before the class II T~Tauri phase. It is unlikely that tertiary interactions \citep[e.g.,][]{fabrycky07} are responsible for the majority of tight binaries \citep{moe18}; rather, migration via a circumbinary disk appears to be the dominant pathway \citep{tokovinin20}. \tw A represents both the lowest mass binary ($M_\mathrm{A} = 0.53 \pm 0.01\,M_\odot$) hosting a circumbinary disk and the longest period binary ($P_\mathrm{inner} = 34.879\pm0.001\,$days) before the population of mutual inclinations transitions from entirely coplanar systems to a broad dispersion of mutual inclinations \citep[see Figure 14,][]{czekala19}. Though the range of binary periods over which this transition occurs is not yet well defined, recent observations have shed light on the dispersion of mutual inclinations at slightly longer binary periods of several months. The recent measurements of WW~Cha by VLTI/GRAVITY have demonstrated that coplanar, truly low mutual inclination ($\theta < 8\degr$) circumbinary disks can and do still exist around longer period ($P=207\,$days), eccentric ($e=0.45$) binaries \citep{gravity21}. The measurement of WW~Cha's orbital properties is important because it enhances the contrast of the stark transition from coplanar systems to a broad distribution of mutual inclinations. As a rule, all planets and disks orbiting binaries with $P \lesssim 40\,\mathrm{days}$ are coplanar, yet at binary periods of $\sim6-10$ months there already exists both a coplanar system (WW~Cha) and a polar-oriented system \citep[HD~98800B, $P=315\,$days;][]{kennedy19}. Discovering new circumbinary disks with host binary periods near the ``transition region'' ($P = 30 - 300\,$days) and measuring their mutual inclinations will help map out the regimes where migration may deliver a coplanar system. 

Substantial degeneracies remain in the relationship between the inner and outer stellar orbits of \tw. These degeneracies stem from two unknowns: 1) whether the circumbinary disk and inner binary are $i < 90\degr$ or $i > 90\degr$ and 2) which mode of the $(\omega_\mathrm{A}, \Omega_\mathrm{outer}, T_{0,\mathrm{outer}})$ posterior is correct. The inferred mutual inclinations corresponding to the four permutations of these degeneracies are delineated in Table~\ref{tab:mut}, under two different prior assumptions ($\sin(\theta)$ or flat).\footnote{Since the value of $\theta_\mathrm{disk-inner}$ is low, values of $\theta_\mathrm{inner-outer}$ and $\theta_\mathrm{disk-outer}$ are similar in all cases. We will refer to $\theta_\mathrm{inner-outer}$ in what follows but the points apply equally to $\theta_\mathrm{disk-outer}$ as well.} These separate into a coplanar configuration ($\theta_\mathrm{inner-outer} \approx 0\degr$),  two orthogonal configurations ($\theta_\mathrm{inner-outer} \approx 90\degr$) and a retrograde configuration ($\theta_\mathrm{inner-outer} > 90\degr)$. The field population of triples with projected outer separations $< 50\,$au exhibits a high degree of alignment between inner and outer orbital planes \citep[average mutual inclination $< 20\degr$;][]{tokovinin17}. Considering this backdrop, we suspect that the \tw\ Aa--Ab and A--B orbital planes are nearly coplanar as well. We provide further hydrodynamical evidence for this scenario in the next subsection.

\subsection{Disk truncation}
The time-dependent gravitational potential of a binary star will influence the radial extent of a protoplanetary disk---clearing an interior cavity in a circumbinary configuration and truncating the outer edge in a circumstellar configuration. \tw\ is noteworthy because both types of disk truncation are present in the same system. 

Several analytical and numerical works have derived the radius (usually conveyed as a ratio relative to the binary semimajor axis, $r/a$) to which an interior, coplanar, and eccentric binary like Aa--Ab ($e_\mathrm{inner} = 0.63 \pm 0.01$) is predicted to clear the inner edge of circumbinary disk \citep{artymowicz94,miranda15,miranda17,thun17,hirsh20}. Using an edge definition where the density falls to 50\% of its peak value, \citet{artymowicz94} found that  $r/a \sim 2 - 3$, though they also noted that nonaxisymmetric waves at the rim of the disk make it difficult to define the edge location uniquely. \citet{miranda15} studied circumbinary disk truncation across a range of mutual inclinations, finding that the truncation radius was smaller for more misaligned systems. For coplanar systems, their results agreed with \citet{artymowicz94}. Using a suite of numerical simulations, \citet{miranda17} found that the truncation radius of coplanar circumbinary disks is $r/a \sim 1.7 - 2.6$ (using a 10\% of peak density definition) though there is ambiguity in both the sharpness of the inner edge and peak location. \citet{thun17} used a 2D grid based setup to derive scalings for various inner edge thresholds. For an $e\approx0.6$ binary, they found that the inner edge scales as  $r/a \sim 3.5$ for a 10\% of peak definition, but found that the 50\% location scales like $r/a \sim 4-6$. More recently, \citet{hirsh20} used a smoothed-particle hydrodynamics (SPH) setup to derive 50\% thresholds for an $e\approx 0.6$ binary, and found  values more in line with the studies by \citet{artymowicz94} and \citet{miranda15}: $r_\mathrm{inner}/a_\mathrm{inner} \approx 3.0 - 3.5$. \citet{hirsh20} noted that \citet{thun17}'s discrepancy with previous results could be attributable to their choice of inner polar grid boundary. For the \tw A circumbinary disk, the best-fit interior cavity radius of $r_\mathrm{cav} / a_\mathrm{inner} \approx 3$ inferred from SED modeling (\S\ref{subsec:sed}) is squarely near the median of the aforementioned theoretical predictions, though solutions with $r_\mathrm{cav} / a_\mathrm{inner} \approx 4.5$ are still consistent with the SED data.

The inner edge of a circumbinary disk is also expected to be eccentric due to resonant interactions between the binary and the disk. For an $e_\mathrm{inner}=0.63$ binary, \citet{thun17} found the inner edge would have $e_\mathrm{disk} \approx 0.4$; \citet{ragusa20} (using SPH simulations of more extreme mass ratio binaries) also found similar results: $e_\mathrm{disk} \approx 0.05 - 0.35$. \citet{munoz20} used linear theory of perturbed, pressure-supported disks to solve for the eccentricity profile and showed that the eccentric modes are concentrated to within $r/a < 2$ and drop off exponentially after $r/a \sim 10$ (corresponding to 1.7\,au for \tw A). In simulations, the disk eccentricity is $e_\mathrm{disk} < 0.05$ after $r/a = 10$ \citep{thun17,ragusa20}. Unfortunately, the scale of the inner rim of the circumbinary disk is below the resolution that can be meaningfully probed by the current ALMA observations. If the outer disk ($r \gtrsim 4\,\mathrm{au}$) were eccentric, it would be readily observable as a strong flux asymmetry between the redshifted and blueshifted sides of the disk \citep{czekala15a}. Such an effect is not seen in these observations.

For equal mass binaries, nearly circular ($e \lesssim 0.1$), coplanar orbits are expected to truncate the outer edge of a circumstellar disk at radii $r_\mathrm{outer} / a_\mathrm{outer} \sim 0.32 - 0.38$ \citep{artymowicz94,miranda15}. More eccentric binaries are more effective at truncating the disk: $r_\mathrm{outer} / a_\mathrm{outer} \sim 0.20 - 0.25$ for $e=0.2-0.4$. Coplanar configurations are also the most effective at truncating the disk: as the mutual inclination $\theta_\mathrm{inner-outer}$ increases, $r_\mathrm{outer}$ also increases \citep[about 20\% larger for $\theta_\mathrm{inner-outer} = 90\degr$ and about 40\% larger for $\theta_\mathrm{inner-outer} = 135\degr$; see Figure 4 of][]{miranda15}. Using the 95\% enclosed mass limit of $8.5\pm 0.2$\,au from the CO modeling and our orbital constraint of $a_\mathrm{outer} = 63\pm18\,$au means that $r_\mathrm{outer}/a_\mathrm{outer} \approx 0.10 - 0.19$ for \tw. 

\begin{figure*}[t]
\begin{center}
\includegraphics[width=0.9\linewidth]{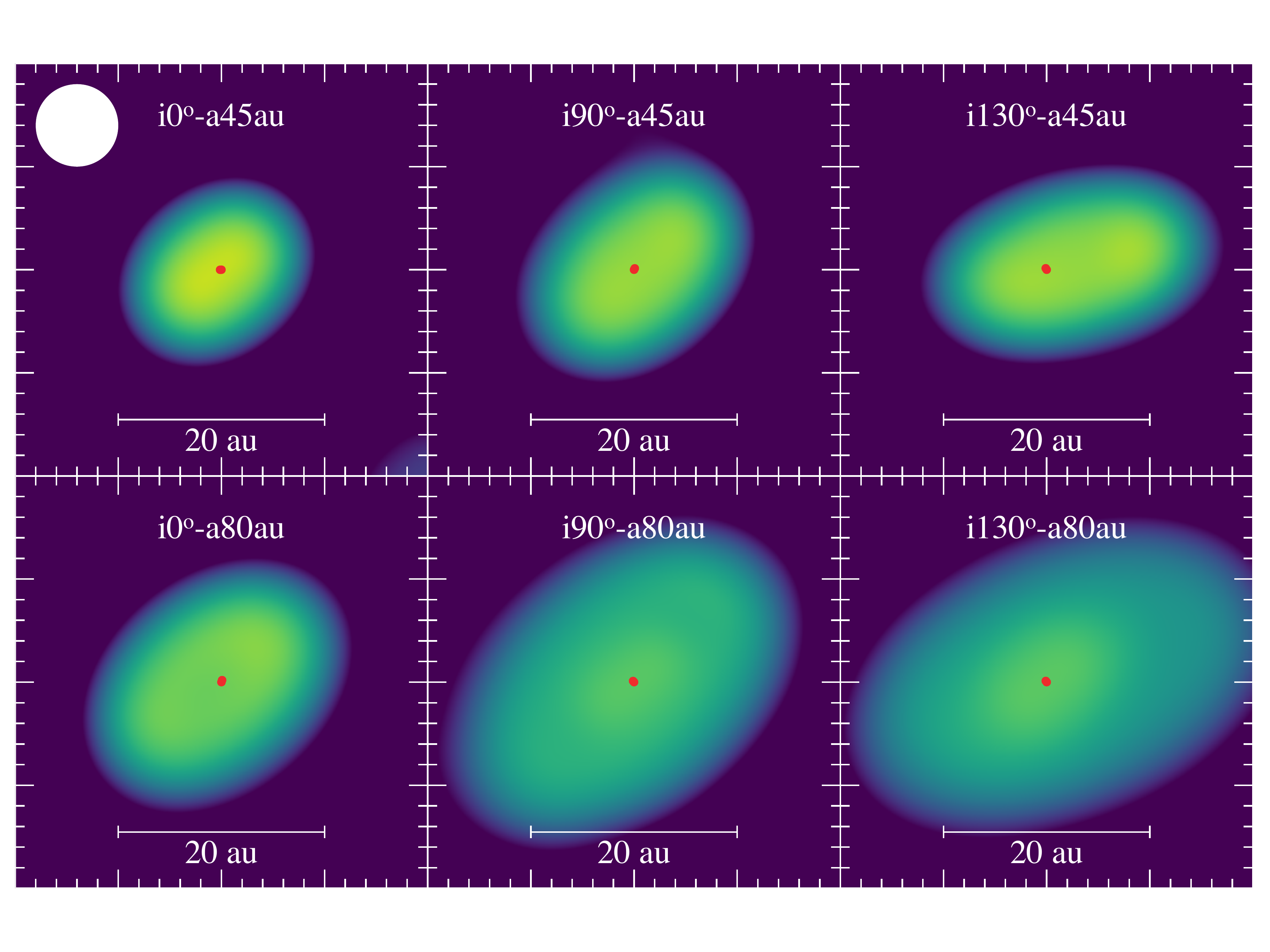}
\figcaption{Gas surface density in SPH simulations of the \tw A binary, circumbinary disk, and exterior companion \tw B, over a range of $\theta_\mathrm{inner-outer}$ mutual inclinations (\texttt{i}) and semi-major axes $a_\mathrm{outer}$ (\texttt{a}). The red dots at the center of the circumbinary disk represent the positions of the stars Aa and Ab (on top of each other) after 5 orbits of the outer companion B (not shown in frame). The images are convolved with an 8 au Gaussian beam (white circle in the top left) for better comparison with the observations shown in Fig.~\ref{fig:moments}. The orbit of B which most closely reproduces the outer radius of the circumbinary disk corresponds to the coplanar one with the smaller semi-major axis ($\theta_\mathrm{inner-outer} = 0\degr$, $a_\mathrm{outer} = 45\,$au; \texttt{i0$\degr$-a45}). Each representation is shown from the perspective of an Earth observer.
\label{fig:hydro}}
\end{center}
\end{figure*}

To further investigate the role of external disk truncation in the \tw\ system, we performed 3D SPH simulations of the interactions between the \tw A binary, circumbinary disk, and \tw B using the \textsc{Phantom} code \citep{phantom}. The SPH method is well suited for misaligned disk simulations given that there is no preferred geometry and angular momentum is conserved to the accuracy of the time-stepping scheme (see e.g. \citealt{price12}). We used $10^6$ gas particles to model the circumbinary disc and set the initial inner and outer radii of the disk to $r_{\rm in}=2$ au and $r_{\rm out}=20$ au, respectively. The surface density initially followed a power-law profile ($\Sigma \propto R^{-1}$) and the temperature profile followed the power law profile $T = 34\,\mathrm{K}\,(r/10\,\mathrm{au})^{-0.5}$, as in \citet{price18}. The disc total mass was set to $0.01\,M_\odot$, which allowed us to neglect the disk self-gravity. Furthermore, we assumed that the disc is locally isothermal, where the sound speed follows a power-law $c_{\rm s} \propto r^{-1/4}$ with $H/r=0.05$ at $r=10$ au. Finally, we adopted a mean Shakura-Sunyaev disc viscosity $\alpha_{\rm SS}\approx5\times10^{-3}$. 

We included all three stars in the simulation (\tw Aa, Ab, and B, each represented by a sink particle), which interact with the gas via gravity and accretion \citep{bate95}. The accretion radii of Aa and Ab were set equal to $0.01$ au, while the accretion radius of B was set equal to 10\% of the Hill radius. These values ensure that the inner and outer regions of the disk are properly modeled while keeping computational costs reasonable (see e.g. \citealt{price18}, \citealt{cuello19} and \citealt{menard20}). 

The masses and orbits of stars Aa and Ab were initialized to the best-fit values listed in Table~\ref{tab:orbits}. The circumbinary disc around Aa and Ab was initially in the same plane as the Aa--Ab binary orbit. The outer companion B was set on a circular ($e_{\rm outer}=0$) orbit with semi-major axis $a_{\rm outer}$ and mutual inclination with respect to the binary orbital plane of $\theta_{\rm inner-outer}$. In principle, the stellar orbits are allowed to change as mass is accreted, but since an insignificant amount of material was accreted throughout the simulation, the change in stellar orbits was negligible. We ran six simulations corresponding to the coplanar, orthogonal, and retrograde orbits of $\theta_\mathrm{inner-outer} = [0\degr, 90\degr, 130\degr]$ and two different values of $a_\mathrm{outer}=[45\,\mathrm{au}, 80\,\mathrm{au}]$. We used $M_\mathrm{B} = 0.4\,M_\odot$ and $e_\mathrm{outer} = 0$ for all simulations. A visualization of the results (convolved with an 8\,au beam) is provided in Figure~\ref{fig:hydro}, showing that the outer edge of the disk is sensitive to choices of $\theta_\mathrm{inner-outer}$ and $a_\mathrm{outer}$. The smallest disks were produced in the coplanar simulations ($\theta_\mathrm{inner-outer} = 0\degr$). The simulation parameters corresponding to the $i_\mathrm{disk} > 90\degr$ and $v_\mathrm{B} \uparrow$ scenario ($\theta_\mathrm{inner-outer} = 0\degr$, $a=45\,$au, $e=0$:  \texttt{i0$\degr$-a45au}) deliver an outer disk edge (10\,au) that most closely matches the ALMA observations. This suggests that the true semi-major axis of the A--B binary lies closer to the lower range of its estimate ($a_\mathrm{outer}=63\pm18\,$au), the true eccentricity of the A--B binary is significantly non-zero ($e_\mathrm{outer}=0.3\pm0.2$), or both. 

\subsection{Comparison to other systems}
The HD~98800 multiple system, coincidentally in the same TW Hydra association as \tw\ (and thus, similarly aged at $\sim 10$\,Myr), is an apt analog system to \tw. Because the HD~98800B circumbinary disk is in a hierarchical multiple system, it also experiences both interior and exterior disk truncation forces (from the Ba--Bb binary and from the wider A companion, respectively). \citet{kennedy19} spatially resolved the circumbinary disk with ALMA and convincingly demonstrated that it is in a circumpolar ($\theta_\mathrm{disk-inner} \approx 90\degr$) configuration. \citet{franchini19} demonstrated that the relatively small cavity size ($r_\mathrm{cav}/a_\mathrm{inner} \sim 1.5 - 2.5$) is a consequence of the reduced torques from an orthogonally-oriented binary \citep[see also][]{miranda15}.

The exterior companion HD~98800A (which is itself a spectroscopic binary Aa--Ab, but here is effectively treated as a single star) orbits with $a_\mathrm{outer} = 54$\,au, $e=0.52$ and $\theta_\mathrm{outer} \approx 65\degr$.\footnote{\citet{kennedy19} were unable to break the $i_\mathrm{disk} < 90\degr$ or $i_\mathrm{disk} > 90\degr$ degeneracy, but the computed value of $\theta_\mathrm{outer}$ is similar for both cases.} The HD~98800B disk outer edge is $\lesssim 7$\,au, suggesting that the eccentric outer companion is much more effective at outer disk truncation ($r_\mathrm{outer}/a_\mathrm{outer} \sim 0.13$), even though the disk and the outer companion are substantially misaligned. The disk may have survived as long as it has due to the combined effect of the inner binary stopping the accretion onto the central source \citep{kuruwita19} and the outer companion(s) stopping the viscous spreading of the disk \citep{ribas2018}. 

The HD~100453 multiple system is also a useful reference point for exterior disk truncation with a misaligned companion. The primary star HD~100453A \citep[$1.7\,M_\sun$ A9Ve;][]{dominik03} is surrounded by a disk whose 1.4\,mm continuum emission extends to $\approx 40\,$au and whose CO emission extends to $\approx 100$\,au \citep{wagner18,vanderplas19}. Scattered light observations of the disk revealed spiral arms \citep{wagner15} and narrow-lane shadows \citep{benisty17}. HD~100453B is an external $0.2\,M_\sun$ companion located at $\approx 100$\,au projected distance \citep{chen06,collins09}. Definitive conclusions about the influence of B on the disk are made difficult by the uncertainty in its orbit; however, recent analysis by \citet{gonzalez20} supports a scenario where the disk and binary plane are substantially misaligned (60\degr), since coplanar orbits consistent with the astrometric data would be otherwise inconsistent with the disk morphology, including the spirals and observed velocity field. The spiral features and narrow lane shadows seen in scattered light also suggest a complicated inner disk structure induced by an undetected, substellar companion interior to the disk \citep{vanderplas19,rosotti20,nealon20}. The misalignments in this potentially triple system can be explained as being driven by the outer B, which drives the outer disk, substellar companion, and inner disk to precess and occasionally undergo Kozai-Lidov oscillations \citep{nealon20}.

The LTT~1445ABC triple system, which hosts a transiting exoplanet \citep{winters19}, also bears mentioning in the context of \tw. LTT~1445ABC consists of three mid to late M dwarfs in a hierarchical configuration: B--C forms the inner binary and A is an outer tertiary and the most massive star in the system. The planet transits A; the entire stellar system is co-planar. The \tw\ system is something of a pre-main sequence counterpoint to LTT~1445ABC, in particular the configuration of its circumbinary disk contrasts with the fact that in LTT~1445A the planet transits the single star. Though if the \tw Aa-Ab stars were considered together, \tw A would also be the primary star in the system.

\section{Conclusions}
\label{sec:conclusions} 
Our main conclusions from this study of the \tw\ system are as follows. 

\begin{itemize}
    \item We detected \twelve\ $J=2-1$ and \thirteen\ $J=2-1$ emission from the \tw A circumbinary disk for the first time.
    \item We forward modeled the \twelve\ $J=2-1$ visibilities to derive an updated disk orientation ($\Omega_\mathrm{disk} = 116.5\pm0.4$ and $i_\mathrm{disk}=48.8\pm0.7$) and infer the stellar mass enclosed by the disk $M_\mathrm{A} = M_\mathrm{Aa} + M_\mathrm{Ab} = 0.534\pm 0.010$.
    \item We combined the disk dynamical constraints with extant radial velocity and astrometric measurements of \tw Aa, Ab, and B to infer individual stellar masses $0.29\pm0.01\,M_\odot$, $0.24\pm0.01\,M_\odot$, and $0.40\pm0.28\,M_\odot$, respectively. 
    \item We drew constraints on the orbital architecture of the system, and inferred that the plane of the inner Aa--Ab binary and its circumbinary disk are coplanar (misalignment $<6\degr$ with 68\% confidence). There are several degenerate solutions for the mutual inclination between the orbital planes of the inner (Aa--Ab) and outer (A--B) stellar orbits, however, SPH simulations lend support to the coplanar solution. 
    \item We found the inner and outer radii of the circumbinary disk ($0.50-0.75\,$au and $8.5\pm0.2\,$au, respectively) to be consistent with theoretical predictions of dynamical truncation from coplanar orbits.
\end{itemize}

\acknowledgments
IC would like to thank Rob De Rosa and Eric Nielsen for help with questions about orbital inference; Brian Mason for assistance with the Washington Double Star Catalog \citep{mason01}; Ryan Loomis for assistance with self-calibration; and Christophe Pinte for assistance with MCFOST. IC and AR would like to thank Elise Furlan for providing the calibrated IRS/\emph{Spitzer} spectrum of TWA~3.
IC was supported by NASA through the NASA Hubble Fellowship grant HST-HF2-51405.001-A awarded by the Space Telescope Science Institute, which is operated by the Association of Universities for Research in Astronomy, Inc., for NASA, under contract NAS5-26555.
This project has received funding from the European Union's Horizon 2020 research and innovation programme under the Marie Sk\l{}odowska-Curie grant agreement No 210021.
EC acknowledges NASA grants 80NSSC19K0506 and NNX15AD95G/NEXSS. 
The National Radio Astronomy Observatory is a facility of the National Science Foundation operated under cooperative agreement by Associated Universities, Inc.
This paper makes use of the following ALMA data: ADS/JAO.ALMA\#2018.1.01545.S.
ALMA is a partnership of ESO (representing its member states), NSF (USA) and NINS (Japan), together with NRC (Canada), MOST and ASIAA (Taiwan), and KASI (Republic of Korea), in cooperation with the Republic of Chile.
The Joint ALMA Observatory is operated by ESO, AUI/NRAO and NAOJ. This research has made use of NASA's Astrophysics Data System Bibliographic Services.

\software{CASA \citep[v4.4;][]{mcmullin07}, DiskJockey \citep{czekala15a,disk_jockey_zenodo}, RADMC-3D \citep{dullemond12}, emcee \citep{foreman-mackey13}, Astropy \citep{astropy13}, PyMC3 \citep{pymc3}, Theano \citep{theano}, MCFOST \citep{mcfost}, {\sc Phantom} \citep{phantom}, {\sc splash} \citep{splash}, uvplot \citep{uvplot_mtazzari}}.

\clearpage

\bibliographystyle{aasjournal.bst}
\bibliography{biblio.bib}

\appendix 

\section{\thirteen\ $J=2-1$ channel maps}
\label{sec:appendix}

\begin{figure*}[h]
\begin{center}
\includegraphics{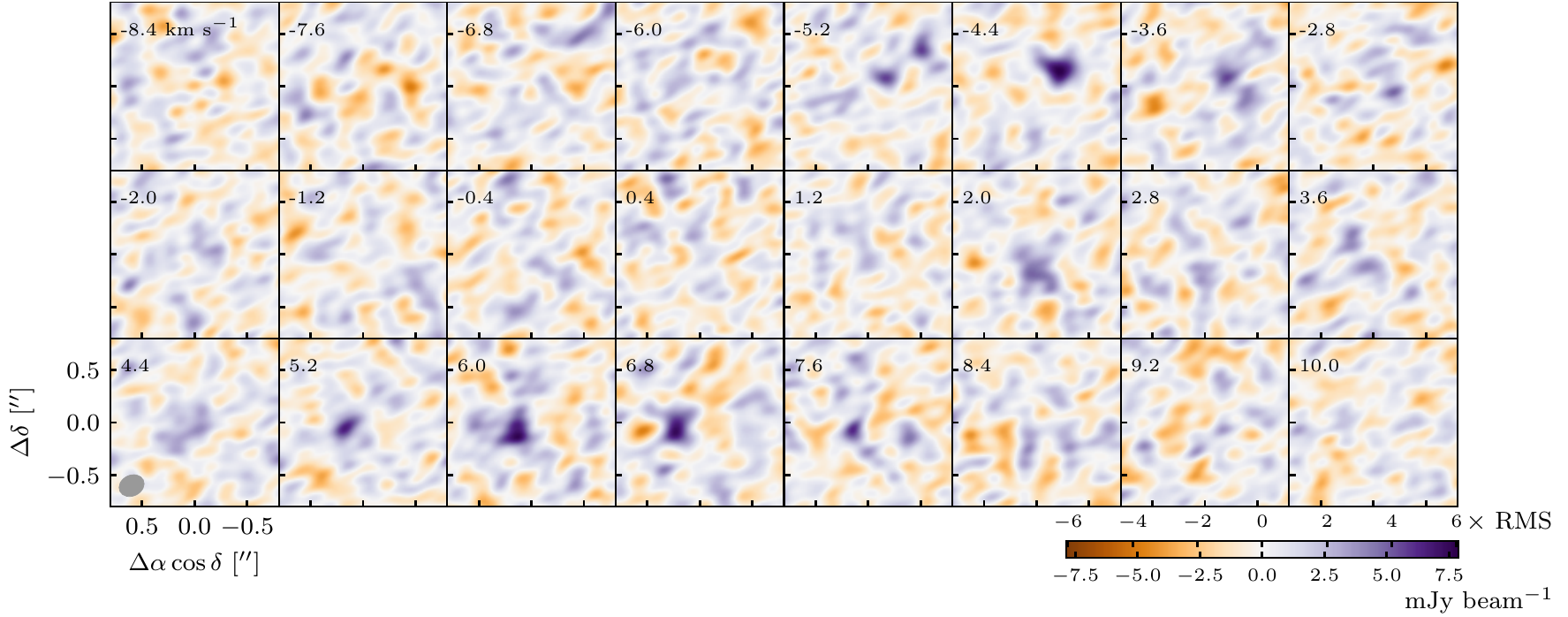}
\figcaption{Continuum-subtracted \thirteen\ $J=2-1$ data channel maps. Velocity scale is labeled in the LSRK frame.
\label{fig:chmaps_13}}
\end{center}
\end{figure*}

\end{document}

%% file: table.tex
\begin{deluxetable}{lcc}
\tablecaption{Orbital parameters \label{tab:orbits}}
\tablehead{\colhead{Parameter} & \colhead{primary solution} & \colhead{alternate solution}}
\startdata
\cutinhead{Sampled}
$P_\mathrm{inner}$ [days] & $34.879 \pm 0.001$ & $34.879 \pm 0.001$\\
$a_\mathrm{inner}$ [mas] & $4.63 \pm 0.04$ & $4.63 \pm 0.04$\\
$M_\mathrm{Ab}$ [$M_\odot$] & $0.24 \pm 0.01$ & $0.24 \pm 0.01$\\
$e_\mathrm{inner}$ & $0.63 \pm 0.01$ & $0.63 \pm 0.01$\\
$i_\mathrm{inner}$ [\degr] & $131.5 \pm 0.8$ & $48.5 \pm 0.8$\\
$\omega_\mathrm{Aa}$\tablenotemark{a} [\degr] & $81 \pm 1$ & $81 \pm 1$\\
$\Omega_\mathrm{inner}$\tablenotemark{b} [\degr] & $104 \pm 9$ & $112 \pm 9$\\
$T_{0,\mathrm{inner}}$ [JD - 2,450,000] & $2704.57 \pm 0.07$ & $2704.57 \pm 0.07$\\
$P_\mathrm{outer}$ [yrs] & $548 \pm 244$ & $555 \pm 249$\\
$M_\mathrm{B}$ [$M_\odot$] & $0.41 \pm 0.28$ & $0.40 \pm 0.28$\\
$e_\mathrm{outer}$ & $0.3 \pm 0.2$ & $0.3 \pm 0.2$\\
$i_\mathrm{outer}$ [\degr] & $139 \pm 13$ & $139 \pm 13$\\
$\omega_\mathrm{A}$\tablenotemark{a} [\degr]& \nodata\tablenotemark{c} & \nodata\tablenotemark{c}\\
$\Omega_\mathrm{outer}$\tablenotemark{b} [\degr]& \nodata\tablenotemark{c} & \nodata\tablenotemark{c}\\
$T_{0,\mathrm{outer}}$ [JD - 2,450,000]& \nodata\tablenotemark{c} & \nodata\tablenotemark{c}\\
$\varpi$ [$\arcsec$] & $27.31 \pm 0.12$ & $27.31 \pm 0.12$\\
$\sigma_\rho$ [$\arcsec$] & $0.009 \pm 0.004$ & $0.009 \pm 0.004$\\
$\sigma_\theta$ [$\degr$] & $0.024 \pm 0.008$ & $0.024 \pm 0.008$\\
$\sigma_\mathrm{CfA}$ [km s${}^{-1}$] & $3.8 \pm 0.3$ & $3.8 \pm 0.3$\\
$\sigma_\mathrm{Keck}$ [km s${}^{-1}$] & $0.8 \pm 0.1$ & $0.8 \pm 0.1$\\
$\sigma_\mathrm{FEROS}$ [km s${}^{-1}$] & $3.4 \pm 0.6$ & $3.4 \pm 0.6$\\
$\sigma_\mathrm{du\;Pont}$ [km s${}^{-1}$] & $2.1 \pm 0.4$ & $2.1 \pm 0.4$\\
(Keck - CfA) [km s${}^{-1}$] & $-1.3 \pm 0.5$ & $-1.3 \pm 0.5$\\
(FEROS - CfA) [km s${}^{-1}$] & $1.3 \pm 1.0$ & $1.3 \pm 1.0$\\
(du Pont - CfA) [km s${}^{-1}$] & $-0.2 \pm 0.7$ & $-0.2 \pm 0.7$\\
\cutinhead{Derived}
$M_\mathrm{Aa}$ [$M_\odot$] & $0.29 \pm 0.01$ & $0.29 \pm 0.01$\\
$M_\mathrm{A}$ [$M_\odot$] & $0.53 \pm 0.01$ & $0.53 \pm 0.01$\\
$a_\mathrm{inner}$ [au] & $0.170 \pm 0.001$ & $0.170 \pm 0.001$\\
$a_\mathrm{outer}$ [au] & $63 \pm 18$ & $64 \pm 19$\\
$r_{p,\mathrm{outer}}$ [au] & $45 \pm 21$ & $45 \pm 21$\\
\enddata
\tablenotetext{a}{The argument of periastron of the primary. $\omega_\mathrm{secondary} = \omega_\mathrm{primary} + \pi$.}
\tablenotetext{b}{The ascending node is identified as the point where the secondary body crosses the sky plane \emph{receding} from the observer.}
\tablenotetext{c}{Posterior is non-Gaussian and not accurately represented by a summary statistic; see Figure~\ref{fig:corner}}
\end{deluxetable}

%% file: mut.tex
\begin{deluxetable}{cccccc}[b]
\tablecaption{Inferred mutual inclinations \label{tab:mut}}
\tablehead{\colhead{$i_\mathrm{disk}$} & \colhead{$v_B$} & \colhead{$p(\theta)$} & \colhead{$\theta_\mathrm{disk-inner}$} & \colhead{$\theta_\mathrm{inner-outer}$} & \colhead{$\theta_\mathrm{disk-outer}$} \\ 
\colhead{$[\degr]$} & \colhead{} & \colhead{} &  \colhead{$[\degr]$} & \colhead{$[\degr]$} & \colhead{$[\degr]$}}
\startdata
$< 90$ & $\uparrow$ & $\sin(\theta)$ & $<8$ & $99_{-16}^{+17}$ & $99_{-17}^{+18}$\\
$< 90$ & $\uparrow$ & flat & $<3$ & $100_{-16}^{+17}$ & $100_{-18}^{+18}$\\
$< 90$ & $\downarrow$ & $\sin(\theta)$ & $<8$ & $140_{-10}^{+10}$ & $143_{-10}^{+7}$\\
$< 90$ & $\downarrow$ & flat & $<3$ & $142_{-11}^{+10}$ & $145_{-10}^{+7}$\\
$> 90$ & $\uparrow$ & $\sin(\theta)$ & $<13$ & $22_{-10}^{+14}$ & $20_{-12}^{+19}$\\
$> 90$ & $\uparrow$ & flat & $<6$ & $<20$ & $<15$\\
$> 90$ & $\downarrow$ & $\sin(\theta)$ & $<13$ & $76_{-17}^{+14}$ & $82_{-16}^{+13}$\\
$> 90$ & $\downarrow$ & flat & $<6$ & $75_{-17}^{+15}$ & $82_{-17}^{+14}$\\
\enddata
\tablecomments{One $\sigma$ asymmetric error bars are reported for all unimodal distributions. Sixty eight percent confidence upper limits are reported for one-sided distributions.}
\end{deluxetable}